# Simultaneous reconstruction of phase and amplitude contrast from a single holographic record

**Tatiana Latychevskaia**[*] **and Hans-Werner Fink**

*Institute of Physics, University of Zurich, Winterthurerstrasse 190, Zurich, CH-8057, Switzerland*
[*]*Corresponding author:* tatiana@physik.uzh.ch

**Abstract:** We present a reconstruction technique for simultaneous retrieval of absorption and phase shifting properties of an object recorded by in-line holography. The routine is experimentally tested by applying it to optical holograms of a pure phase respectively amplitude object of micrometer dimensions that has been machined into a glass-plate using a focused ion beam. The method has also been applied to previously published electron holograms of single DNA molecules.

## 1. Introduction

The holography principle, originally proposed by D. Gabor in 1948 [1], is based on a divergent coherent wave in which path an object is placed that scatters part of the beam elastically. The scattered wave interferes with the un-scattered one and by this forming the hologram in which all information about the object wave, its amplitude and phase, is contained. This conceptually simple scheme of in-line holography has the advantage that no further optical elements such as lenses are required. This is of particular significance when electrons are used since electron lenses exhibit intrinsic aberrations which are associated with a loss of resolution.

The holographic image is usually recorded by a CCD camera followed by numerical reconstruction of the digital record. Most reconstruction methods assume either pure amplitude [2] or pure phase [3-5] objects. However, in reality, nearly all scattering objects affect both amplitude and phase and furthermore, absorption and phase shifting properties of real objects are in general wavelength-dependent. For instance, in high energy electron holography most biological objects exhibit weak absorption properties and are thus almost transparent to the electron beam. Such objects can be visualized by the measurement of the phase shift they introduce to the electron wave [4-6]. However, at low electron energies [7] of about 100 eV, associated with less radiation damage to biological molecules, objects exhibit both, absorption and phase shifting properties. Consequently, the situation must be described by a complex transmission function. For such objects, a truthful hologram reconstructing routine must accurately retrieve absorption and phase properties.

While there are many known reconstruction routines to retrieve the complex field distribution from two or more holograms [8-14], they all rely on imposing assumptions about the object or its environment. We shall show below that absorption as well as phase shifting properties can simultaneously be retrieved from a single holographic record without prior knowledge or assumptions about the object. The routine is based on propagating the field from the hologram plane where the superposition of the unknown object with the known reference wave forms the hologram back to the object plane where a careful mathematical treatment allows separating the two waves.

## 2. Transmission function

In general, the penetration of a wave through a medium is described by a complex transmission function, consisting of two components, amplitude and phase:

$$e^{-a(r)}e^{-i\phi(r)}. \qquad (1)$$

According to Beer's law, $a(r)$ is the distribution of the object's absorption property, $e^{-a(r)} \approx (1-a(r))$ describes how much of the incident beam is not absorbed by the object, $\phi(r)$ is the object's phase distribution, or the phase delay introduced by the object into an incident beam; $r = (x,y,z)$ represents a point in the object. For a pure absorbing (also called pure amplitude) object its transmission function is $e^{-a(r)}$, and the transmission function of a pure phase object is $e^{-i\phi(r)}$.

In general, the absorption and phase shifting distributions can be written as: $a(r) = a_0 + a_1(r)$, $\phi(r) = \phi_0 + \phi_1(r)$, where $a_0$ and $\phi_0$ are absorption and phase shift of the supporting medium. The transmission function is then:

$$e^{-a_0}e^{-i\phi_0}e^{-a_1(r)}e^{-i\phi_1(r)} = C_0 e^{-a_1(r)}e^{-i\phi_1(r)}, \qquad (2)$$

where $C_0$ is a complex constant describing the transmission of the supporting medium, $C_0 = 1$ for vacuum.

On the other hand, the transmission function can be represented as

$$C_0 e^{-a_1(r)}e^{-i\phi_1(r)} = C_0(1+t(r)) \qquad (3)$$

where the 1 in the brackets on the right side corresponds to the transmittance as if there were no object, and the complex function $t(r)$ is the perturbation caused by the presence of the object. Such representation of the object transmittance as $(1+t(r))$ is useful to provide a simple separation between the reference wave (corresponding to 1) and the object wave (corresponding to $t(r)$).

When an object is illuminated by a spherical divergent wave originating from a point source - $A\dfrac{e^{ikr}}{r}$, the field distribution past the object is:

$$A\frac{e^{ikr}}{r}C_0(1+t(r)) = A\frac{e^{ikr}}{r}C_0 + A\frac{e^{ikr}}{r}C_0 t(r), \qquad (4)$$

where the first term on the right side represents the wave which, when propagating towards the screen, provides the reference wave:

$$A\frac{e^{ikr}}{r}C_0 \rightarrow AC_0 \frac{e^{ikr_s}}{r_s} = AC_0 R, \qquad (5)$$

where $r_s = (x_s, y_s, z_s)$ defines a point on the screen. The second term in Eq.(4) describes the object wave. The total field on the screen is the sum of reference and object waves:

$$A\frac{e^{ikr}}{r}C_0(1+t(\mathbf{r})) \to AC_0(R+O). \tag{6}$$

The transmission of the recorded hologram is:

$$H = |AC_0|^2 |R+O|^2. \tag{7}$$

## 3. Hologram normalization

Prior to the reconstruction procedure, the hologram is normalized in order to eliminate effects of background imperfections. The background image is recorded under the same experimental conditions as the hologram just without the object being present:

$$B = |AC_0|^2 |R|^2. \tag{8}$$

Normalization of the hologram is done by dividing the hologram image by the background image. The resulting image H/B does not depend on the constant coefficient $AC_0$ present in both, reference and object wave. The term $AC_0$ is due to experimental factors like the transmission of the sample supporting medium, the point source intensity or detector sensitivity; that is to all experimental elements in the set-up that influence the reference and the object wave and the detection of their superposition in the same manner. By removing these factors we end up with a unified hologram reconstruction routine which is applicable independent of the details of the experimental conditions.

The second step following normalization is to subtract 1 for the holographic image to become a function oscillating around zero. Since the surrounding hologram area is an infinite area filled with zeros, a cosine-window filter is applied to smoothly bring the edges of the holographic record down to zero in order to avoid artifacts otherwise caused by the Fourier-transformations of the hologram.

Altogether, the transmission function of the normalized hologram, in the weak object wave approximation, becomes:

$$\frac{H}{B} - 1 = \frac{R^*O + RO^*}{|R|^2}. \tag{9}$$

## 4. Hologram reconstruction, absorption and phase retrieval

The experimental reconstruction is achieved by illuminating the photo plate hologram with the reference wave, the numerical reconstruction of digital holograms consists of multiplication of the hologram with the reference wave $R = e^{ikr_s}/r_s$ followed by back propagation to the object plane which boils down to evaluating the Kirchhoff-Helmholtz integral transformation [15]:

$$U(x,y,z) \approx -\frac{i}{\lambda z_s^2}\iint H(\mathbf{r}_s)\exp(ik\mathbf{r}\mathbf{r}_s/r_s)d\sigma_s \tag{10}$$

where integration is performed over the screen surface. The transformation in emission vector coordinates is given by :

$$U(x, y, z) = -\frac{i}{\lambda} \int\int H(\kappa) \exp[ikz(1 - \kappa_x^2 - \kappa_y^2)] \exp[ik(x\kappa_x + y\kappa_y)] d\kappa_x d\kappa_y \quad (11)$$

with the emission vector $\kappa = \frac{1}{r_s}(x_s, y_s, z_s)$. The initial step of the hologram reconstruction is a transformation from Cartesian screen coordinates to $\kappa$-coordinates which projects the hologram onto a spherical surface. Eq.(11) can then be employed to directly provide the complex field distribution at $r = (x, y, z)$ by a single Fourier transform. However, a single Fourier transform requires the correct sampling of the spherical wave term in Eq.(11). Instead, we use two steps in the reconstruction routine: first, we calculate the field distribution in the plane $z_0 = 0$:

$$U(x_0, y_0, 0) = -\frac{i}{\lambda} \int\int H(\kappa) \exp[ik(x_0\kappa_x + y_0\kappa_y)] d\kappa_x d\kappa_y \quad (12)$$

which is simply the back Fourier-transform of the holographic image.

Following that, the field is propagated from $z_0 = 0$ to some object's plane $z$ which is again calculated by the Kirchhoff-Helmholtz integral:

$$U(x, y, z) = \frac{i}{\lambda} \int\int U(x_0, y_0, 0) \frac{\exp[ik|\vec{r} - \vec{r_0}|]}{|\vec{r} - \vec{r_0}|} dx_0 dy_0. \quad (13)$$

Using the approximation $|\vec{r} - \vec{r_0}| \approx r - \frac{r_0 r}{z} + \frac{r_0^2}{2z}$ it follows

$$U(x, y, z) \approx \frac{i}{\lambda z} \exp(ikr) \int\int U(x_0, y_0, 0) \exp\left[-\frac{i2\pi}{z}(x_0 x + y_0 y)\right] \exp\left[\frac{i\pi}{\lambda z}(x_0^2 + y_0^2)\right] dx_0 dy_0 \quad (14)$$

The second step consists of multiplication with a complex spherical wave factor, followed by a forward Fourier transformation. When analytical integration is replaced by numerically integration, the total multiplication factor for the Kirchhof-Helmholz transformation, Eq.(10) becomes $\frac{\Delta_0^2 \Delta_k^2}{\lambda^2} = \frac{1}{N^2}$, where $\Delta_k$ is the sampling size in $k$-space and $\Delta_0$ the sampling size in the $z_0 = 0$ plane. Fast Fourier Transformation provides the definition: $\frac{2\pi \Delta_0 \Delta_k nm}{\lambda} = \frac{2\pi nm}{N}$.

The resulting field is a complex function including also the incident wave. To extract the object's transmission function, the result of Eq.(14) must be divided by the incident wave to reveal: $t(r) = re^{-ikr}U(r)$.

While the reconstructed term $t(x, y)$ alone does not directly provide the absorption and phase properties of the object, the sum $(1 + t(x, y))$ provides a simple connection to the absorption and phase properties of the object as expressed by Eq.(3). The object absorption $a(x, y)$ is given by the absolute value of $(1 + t(x, y))$:

$$|1+t(x,y)| = e^{-a(x,y)} \approx (1-a(x,y)). \tag{15}$$

The phase shift $\phi(x,y)$ introduced by the object is extracted as the phase of $1+t(x,y)$:

$$\phi(x,y) = \text{phase}\{1+t(x,y)\}. \tag{16}$$

## 5. Experimental holograms

To verify our simultaneous phase and amplitude reconstruction method, we recorded holograms of patterns engraved into a 0.15 mm thick microscope cover slip as shown in Fig.1. A 3-5 nm thin Ti layer was sputtered onto the glass surface prior to its structuring by using a focused gallium ion beam (FIB) machine. A pure amplitude object as illustrated in Fig.1a was created by depositing a 15 nm thick layer of Au followed by writing the letter $\Psi$ by means of the focused ion beam to locally remove the gold and titanium layer. For creating a pure phase object no absorbing gold was deposited, instead the transparent Ti layer as well as the glass underneath was removed wherever the focused ion beam impinged onto the surface.

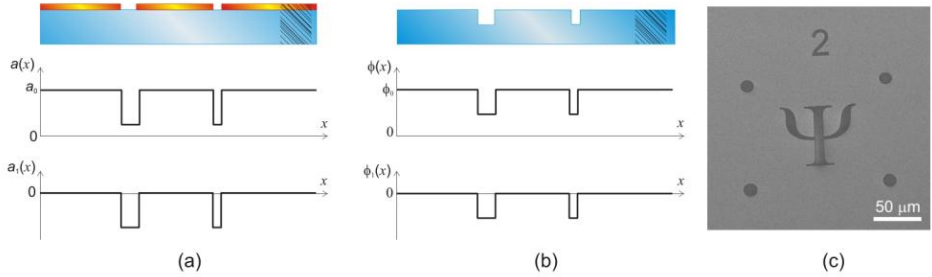

Fig. 1. An illustration to the complex transmission function of the glass samples. (a) A pure absorbing object. (b) A pure phase shifting object. The reference wave is formed by the wave passing through the dashed area, where transmission is $C_0 = e^{-a_0} e^{-i\phi_0}$. (c) Scanning electron image showing the realization of such structures milled with a focused ion beam.

Optical in-line holograms are recorded using green laser light of 532 nm wavelength. A divergent beam is formed by focusing the laser beam onto a pinhole of 20 μm diameter beyond which the object is positioned by a xyz-movable stage, as schematically illustrated in Fig.2. The hologram is recorded on a screen placed at an 80 cm distance from the source and a CCD camera captures the holographic record. Background reference images are recorded using positions of the glass cover slip where no patterns have been engraved.

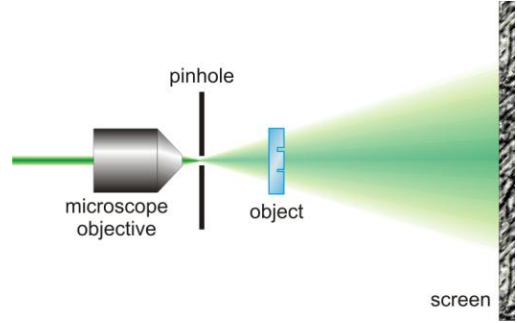

Fig. 2. Experimental arrangement for recording optical in-line holograms. For reasons of presentation, sizes and distances are scaled differently.

## 6. Simultaneous recording of a pure amplitude and a pure phase objects

Fig.3(a) shows hologram of the cover slip with two $\Psi$ letters engraved, one being a purely absorbing (bottom left corner) and the other representing a pure phase object (upper right corner). The hologram size on the screen amounts to 300x300 mm$^2$, the source-object distance is about 1.5 mm. From this single hologram two distributions are reconstructed. The absorption is shown in Fig.3(b) and the phase in Fig.3(c). The pure absorbing $\Psi$ letter is clearly reconstructed in the absorption distribution and seen as darker image because of less absorption in the parts where the gold is removed while the phase $\Psi$ letter remains invisible. In the phase reconstruction, the amplitude $\Psi$ letter appears out of focus. The phase $\Psi$ letter is only visible on the reconstructed phase distribution and appears dark since the phase shift is less where the glass was locally removed by the ion beam.

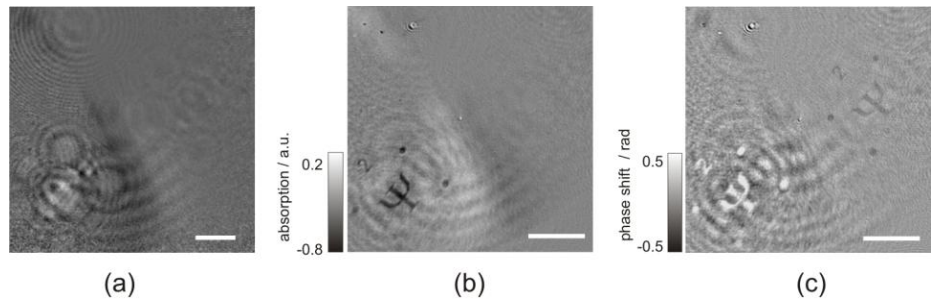

Fig. 3. (a) Normalized hologram. The scale bar corresponds to 50 mm on the screen. (b) Reconstructed absorption distribution. (c) Reconstructed phase distribution. The scale bars in (b) and (c) are about 100μm.

Holograms of the two $\Psi$ letters were also recorded at shorter source-sample distance to provide for larger magnification. Their reconstructions are discussed below in some more detail.

*6.1 Pure absorbing object*

Fig.4(a) shows a hologram of the pure amplitude object. The hologram size is again 300x300 mm$^2$ on the 80 cm distant screen while the source-object distance is now reduced to 0.73 mm to provide higher magnification. The reconstructed absorption distribution $a_1(x, y)$ is displayed in Fig.4(b) and the reconstructed phase distribution $\phi_1(x, y)$ in Fig.4(c). The $\Psi$ letter again appears dark as expected due to less absorption where the ion beam has removed the gold. The pure amplitude $\Psi$ letter shows also up in the reconstructed phase distribution due to the change of the refractive index between the gold plus titanium layers and the glass surfaces.

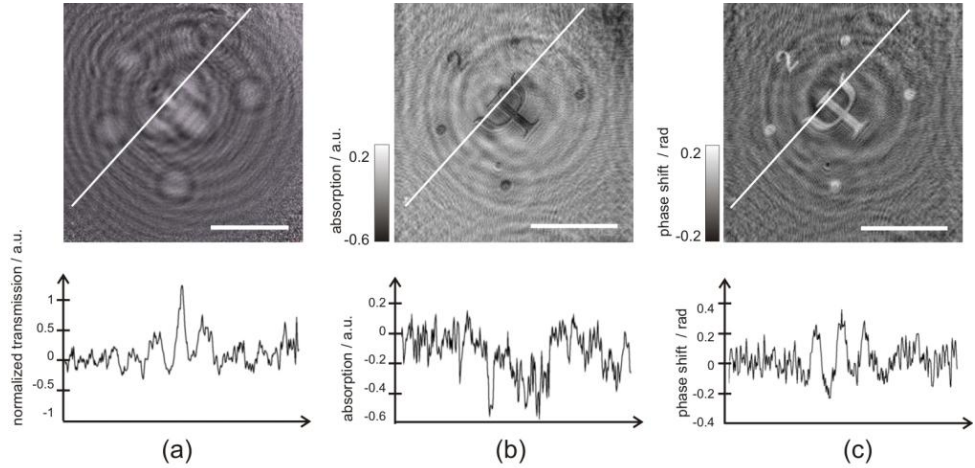

Fig. 4. Normalized hologram of pure amplitude object $\Psi$ letter. The scale corresponds to 100 mm on the screen. (b) Reconstructed absorption distribution. (c) Reconstructed phase distribution. The scale bars in (b) and (c) correspond to 100 µm. The intensity profiles along the white lines are shown below the images.

*6.2 Pure phase object*

A hologram of the pure phase object is shown in Fig.5(a) together with its reconstructed amplitude distribution $a_1(x, y)$ in Fig.5(b) and the reconstructed phase distribution $\phi_1(x, y)$ in Fig. 5(c).

The $\Psi$ letter appears extremely weak in the amplitude reconstruction which may be due to some gallium implantation into the glass. In the phase reconstruction, the maximal phase shift amounts to 0.6 radian. The good contrast indicates that weak phase objects can readily be detected by holographic recording.

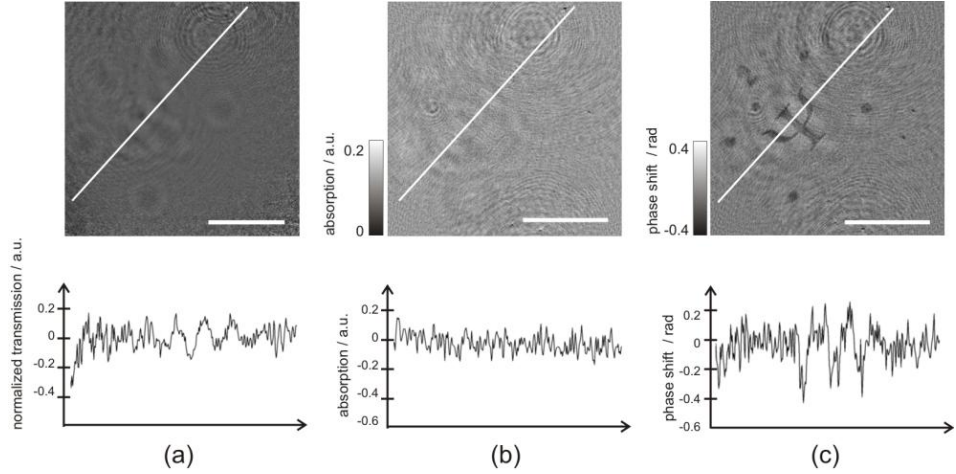

Fig. 5. (a) Normalized hologram of a pure phase object $\Psi$ letter. The scale bar corresponds to 100 mm on the screen. (b) Reconstructed absorption distribution. (c) Reconstructed phase distribution. The scale bars in (b) and (c) correspond to 100 μm. The intensity profiles along the white lines are shown below the images.

## 6. Reconstruction of electron hologram revisited

Finally, we would like to demonstrate the simultaneous amplitude and phase reconstruction to electron holograms which have already been published in the literature [16]. In the original work, electron holograms of DNA and their amplitude reconstructions are shown. Here, in addition to the already published amplitude reconstruction, we retrieve the absorption and phase reconstruction from the same hologram using our method described above. The hologram is created by simply using a screen-capture from the PDF-file of the article. The hologram of a single DNA was taken with 70 eV kinetic energy electrons, captured on a 40 mm diameter detector placed 10cm away from the electron source. The coherent electron source was positioned at 400 nm from the DNA molecule.

The DNA hologram and its amplitude reconstruction are shown in Fig.6., whereby the amplitude distribution is computed as $|t(x,y)|$. The absorption and phase distributions reconstructed from the same hologram are displayed in Fig.7. While the absorption distribution looks similar to the amplitude distribution, the phase distribution shows some structure which cannot be seen in either amplitude or absorption distribution. Besides, when reconstructing the phase distribution at different source-sample distance, it shows a step-like transition in color along the DNA molecule ranging from black to white, which could be attributed to the DNA twists, see Fig.7(b)(Media 1). However, even without being certain about actually seeing the DNA twists, we are confident that by the method of simultaneous absorption and phase distributions retrieval, additional information and further details become available.

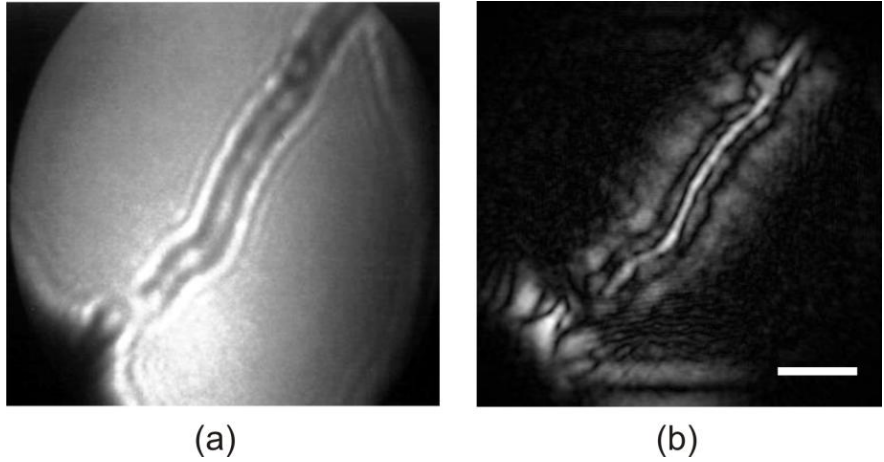

Fig. 6. (a) Screen capture of a hologram of a DNA molecule. The electron energy amounts to 70 eV. (b) Reconstructed amplitude. The scale bar corresponds to 40 nm.

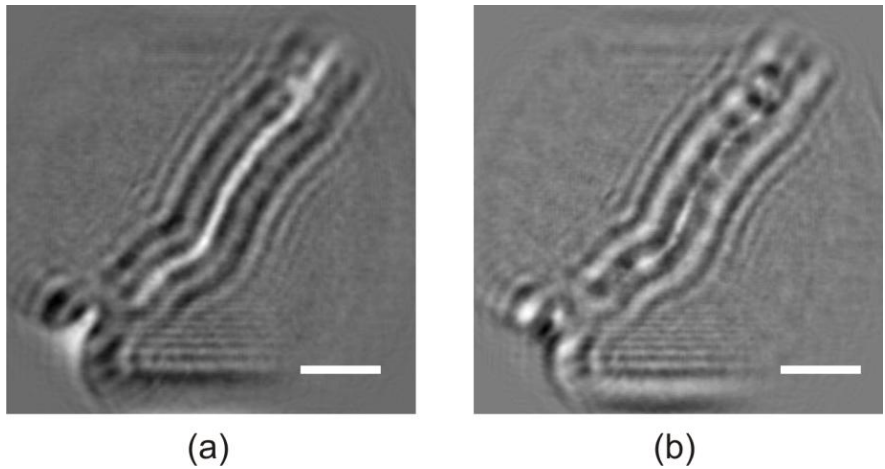

Fig. 7. (a) Reconstructed absorption. (b) Reconstructed phase (Media 1). The scale corresponds to 40 nm.

## 6. Conclusions

A method, free from assumptions of or prior knowledge about a scattering object or its support, has been developed and applied for simultaneous retrieval of absorption as well as phase properties of an object from a single holographic record. Proper normalization renders the reconstruction routine independent of experimental factors, like the transmission of the sample support, source intensity, camera intensity scale, image format etc.

By applying the routine to experimental light optical holograms of micrometer-sized objects we demonstrate that the method is able to retrieve amplitude and phase of the object wave and that high sensitivity even to weak phase objects provides high image contrast.

The method has also been applied to DNA electron holograms, already been published, for retrieval of absorption and phase distributions. In comparison to the conventional reconstruction, being now able to retrieve the phase distribution provides additional structural detail of the just 2nm wide molecule.

Our reconstruction routine is independent of the type of radiation used, being it visible light, x-rays or electrons and free from assumptions or approximations about the object's nature. It can thus be applied to reconstruct high energy electron holograms where biological objects behave as pure phase objects, or low energy electron holograms where both amplitude and phase changes occur in scattering off biological molecules. Whenever aiming at structural biology at the single molecule level by holography, this includes also applications of modern bright pulsed x-ray sources our method shall provide additional sensitivity and with this enhanced structural detail information.

## Acknowledgements

We would like to thank Michael Krüger for machining of the glass sample. The work presented here is supported by the European Project SIBMAR, part of the "New and Emerging Science and Technology" Programme.